\documentclass[twocolumn,amsmath,amstex,eqsecnum,pre,aps]{revtex4-1}
\usepackage{amsmath}
\usepackage{color}
\usepackage{graphicx}
\usepackage{subfigure}
\usepackage{dcolumn}
\usepackage{bm}
\usepackage[colorlinks=true,anchorcolor=blue,
filecolor=green,citecolor=blue,urlcolor=blue,linkcolor=red]{hyperref}

\begin{document}

	\title{Analytical treatment of the structure and thermodynamics of the square-well fluid.}

	\author{Jes\'us Benigno Zepeda-L\'opez$^{1} $, Alexis Torres-Carbajal$^{1} $,Pedro E. Ram\'irez-Gonz\'alez$^{2} $ and Magdaleno Medina-Noyola$^{1} $}

	\affiliation{$^{1} $ Instituto de F\'isica {\sl ``Manuel Sandoval Vallarta"},Universidad Aut\'onoma de San Luis Potos\'i, \'Alvaro Obreg\'on 64, 78000 San Luis Potos\'i, SLP, M\'exico}

	\affiliation{$^{2} $ CONACYT-Instituto de F\'isica {\sl ``Manuel Sandoval Vallarta"},Universidad Aut\'onoma de San Luis Potos\'i, \'Alvaro Obreg\'on 64, 78000 San Luis Potos\'i, SLP, M\'exico}

\begin{abstract}
	The main goal of this work is to accurately reproduce the structural properties of attractive systems modelled by hard-sphere plus square-well (HS+SW) interaction potential. Based on the optimized random phase approximation (ORPA), the attractive part of the interaction potential is treated as a perturbation of the hard-sphere term. We are able to obtain an analytical expression for the structure factor $ S \left( k \right) $ which reproduces the low density limit.
	The microscopical structure of the fluid phase of several SW fluids is computed and compared with Monte Carlo (MC) simulation results showing that the structure factor is well reproduced in a wide range of wave vectors, in addition, the contact and discontinuity values of the radial distribution function are found to be in good agreement. Additionally, we compute the pressure equation of state and perform a quantitative analysis comparing with simulation results found that in a large set of densities and temperatures our approach outperform its linear form. Furthermore, we show that the theoretical approach developed in this study works very well for many thermodynamic states leading us a versatile and confident tool to systematic compute the structure and thermodynamics of SW fluids.
\end{abstract}

\maketitle

\section{\label{SecI}Introduction}
Amorphous solids are very important in a diversity of daily applications \cite{Mort1980,Biroli2011}. Such materials are formed by non-equilibrium processes closely related with glass transition and glassy behaviour \cite{Biroli2011,Angell1995}. Examples of this kind of materials are gels, food and biological matter \cite{Chaves2018,Vintiloiu2008,Hughes2009}. Despite the intense research around this topic, there are many unsolved questions that deserve attention. For instance, the relation between glass-like and gel-like states has been barely studied \cite{Kob2010,Coniglio2014}. Nowadays, is widely accepted that gel transition is a different expression of dynamical arrest behavior \cite{Zaccarelli2005,Zaccarelli2007}. In addition, both, experiments \cite{Gibaud2012} and molecular simulations \cite{Foffi2005} shows that such kind of materials are formed at intermediate densities.

Low-density solids are actually very common in the field of organic solids e.g., protein crystals are usually obtained from a suspension with a few amount of protein. Also, the so-called organic gels are another example of this kind of systems which are commonly used in the food industry \cite{Gibaud2012}. Notwithstanding the experimental data around low-density regimes, in the context of gel-like systems, their relation with the phenomenology of the glass transition is still unclear \cite{Zaccarelli2007,Zaccarelli2005}. Hence, we are interested in such relation and we want to develop general predictions based on microscopical information, which could be useful for interpretation of the physical mechanism that derives in the formation of low-density solids. This goal can be achieved by using the so-called Non-Equilibrium Self-Consistent Generalized Langevin Equation (NE-SCGLE) theory in order to describe the glassy behaviour of the Square-Well model (SW) at low densities.

Recently,  a HS + attractive Yukawa model has been used in the frame-work of the NE-SCGLE theory to study and predict some non-equilibrium properties like ageing in gel-like and spinodal decomposition systems \cite{Olais2015}, this theory requires the microscopical structure of the model as an input. However, is very convenient to extend such study of gel and glass formation using the HS+SW system due to its versatility.	Despite that in the literature there are more sophisticated approximations to determine the microscopical structure or the thermodynamic behaviour of SW fluids, their application over the regions of interest for the afford mentioned problems is a particularly non-trivial task to handle \cite{Brader2006}. On the other hand, an analytical equation for the static structure factor $ S(k) $ also allows us to easily compute and explore different thermodynamic properties  \cite{Hansen} \cite{Cummings1985}.

Since our interest resides in gel-like states, for different purposes, the first step is to develop some confidence about the equilibrium and structural properties. Such properties are the reference that allows to distinguish among equilibrium and non-equilibrium states. Hence, an evaluation of the performance of theoretical calculations of structural properties with simulation results becomes necessary. The main aim of the present work is to make a systematic comparison between the structural properties determined with theoretical basis and Monte Carlo simulations. The final goal is to establish an analytical equation, which accurately describes the structural properties.

The SW fluid is characterised by a pair potential that incorporates a repulsive hard-core interaction and an attractive contribution. This model has been widely used in statistical mechanics in both theoretical approaches \cite{Barker1967,Smith1970,Smith1971,Henderson1976,Henderson1980,Carley1977,Carley1981,Carley1983,DelRio1983,DelRio1985} and computer simulations \cite{Rotenberg1965,Lado1968,Rosenfeld1975,Scarfe1976,Vega1992,Torres2018}. This interaction potential lead us the ability to control, independently, the energy $ \epsilon $ and range $ \lambda $ interaction between molecules \cite{Paschinger2005}. See Eq. \eqref{eq: 01}
\begin{equation}
u(r) =
\left\{
\begin{array}{lcl}
\infty & \hspace*{0.3cm} & r < \sigma \\
- \epsilon & \hspace*{0.3cm} & \sigma \leq r \leq \lambda \sigma \\
0 & \hspace*{0.3cm} &\lambda \sigma < r
\end{array}
\right.
\label{eq: 01}
\end{equation}
The flexibility and features of the potential gives us the opportunity to characterize different liquids and complex fluids \cite{Alejandro1997} and colloidal systems \cite{Duda2009}. Thus, nowadays, the SW fluid is used to gain insight into the thermodynamics, phase and dynamical behaviour of ideal and real fluids \cite{Clare1999,Clare2010}, as well as a model of protein solutions \cite{Valadez2012}. For example, the liquid-phase of the SW fluid is well known and determined by means of perturbation theories \cite{Barker1967,Barker1967_II,Henderson1976}, integral equations \cite{Paschinger2005} and references therein, or simulation techniques \cite{Vortler2000}. In this line of thoughts, the microscopic structure is also a well known studied property, but, for the best of our knowledge, only two analytical equations arise from the literature, one being the Sharma and Sharma (SS) proposal \cite{Sharma1977} and the other being a density expansion from whose analytical expressions are found only up to the first order \cite{Santos}.

The aforementioned schemes are known to be limited to work under certain regimes, e.g., the SS proposal is expected to fail under low temperatures at the small densities regime, as one cannot recover the expected temperature dependency of the structural properties on such conditions. On the other hand, the density expansion scheme has the problem of being accurate just around small values of the density. Due to the small amount of theoretical frameworks in this regard, and their usefulness as an input in theories such as NE-SCGLE, we decided to develop our own approach from which we obtain encouraging results. In consequence, the most important contribution of the present work is a new theoretical approach for the calculation of structural properties of SW fluids, which is found to be consistent with the small density regime as well as having a good overall agreement at higher densities.

Thus, this work is organized as follows. In sec. \ref{SecII} we show phase diagrams of the SW fluid and the thermodynamic states of interest used as a reference for our study. In sec . \ref{SecIII} we detail the procedure used to determine our analytical equation for $ S(k) $. Then, in sec. \ref{SecIV} results for the structure factor and the pair correlation function are analysed. In sec \ref{SecV} a qualitative comparison at the level of the pressure equation is performed between the theoretical approach and simulation results. Finally, in sec. \ref{SecVI} we present some concluding remarks.

\section{\label{SecII}Reference System}
\begin{figure}[htp!]
	\includegraphics[scale=0.49]{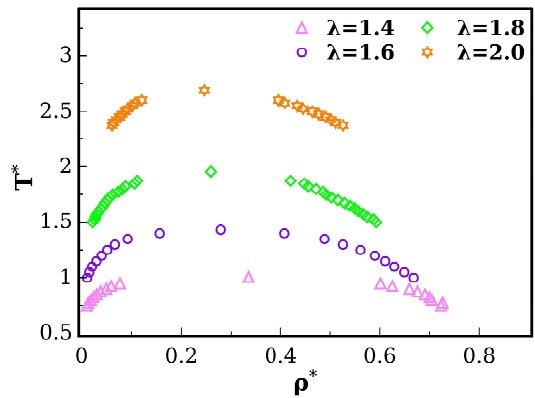}
	\caption{Phase diagram of the Square-Well fluid for different values of $ \lambda $ determine through the Gibbs ensemble Monte Carlo.}
	\label{Fig: 01}
\end{figure}

The relevance of the SW fluid is supported by the vast amount of physical systems that are able to be modelled with such interaction potential. Due to the close relation between equilibrium and non-equilibrium phases we present first our calculation of the equilibrium phase diagram for different values of $ \lambda $. Here, employing the Monte Carlo method in the Gibbs ensemble \cite{Panagiotopoulos1987,Panagiotopoulos1988} we determine the phase diagram of the SW fluid for $ \lambda \in \left[1.4, 2.0\right] $. Fig. \ref{Fig: 01} shows results for selected values of $ \lambda $. It is important to say that our results are in complete agreement with previous studies \cite{Vega1992, Valadez2012}. Then, this information allows to select thermodynamic states where our approach and MC results can be directly compared.

\begin{figure}[htp!]
	\includegraphics[scale=0.49]{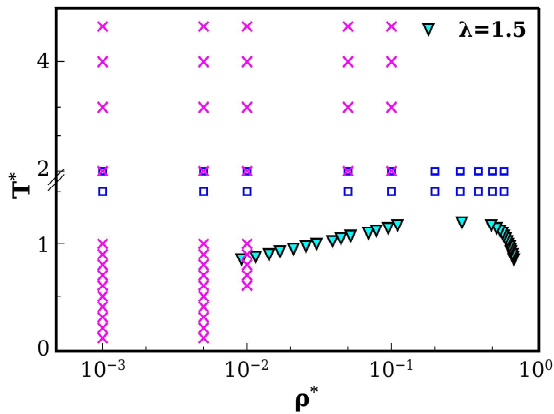}
	\caption{Explored thermodynamic states for the $ \lambda = 1.5 $ SW fluid. Crosses stands for the low density regime where the microscopical structure was analysed. Squares are the thermodynamic states where the pressure was determined in order to perform a qualitative analysis at moderate densities.}
	\label{Fig: 02}
\end{figure}

Is worth to mention that the approach developed and described further below is valid for any $ \lambda $ value, however, in order to focus our discussion in the low density regime, we mainly studied the $ \lambda = 1.5 $ SW fluid whose phase diagram is shown in Fig. \ref{Fig: 02}. Also in the same figure, indicated by crosses, the thermodynamic states where the microscopical structure was determined are signalled. In this work, such thermodynamic states are refereed as the low density regime, where we have used reduced units to express the density as $ \rho^{*} = \rho \sigma^{3} $. Furthermore, with the aim to stablish the extent of our approach a quantity derived from it, as the pressure, is analysed in a wide range of densities by means of the pressure state equation at supercritical temperatures. The thermodynamic states where a comparison between theoretical and simulations results is performed for the pressure are shown in Fig \ref{Fig: 02} as squares. We stress the fact that this analysis covers the low and moderate fluid densities.

\section{\label{SecIII}Analytical equation for S(k) of the Square-Well fluid}

Here, we derive an analytical expression for the static structure factor $S(k)$ performing approximations at the level of the direct correlation function $c(r)$. These functions are directly related through the relationship
\begin{equation}
S(k) = \frac{1}{1 + \rho \hat{c}(k)},
\label{eq: 02}
\end{equation}
where $\rho$ is the system number density and $\hat{c}(k)$ is the Fourier transform of $c(r)$, therefore, the problem becomes the determination of $c(r)$.

The approximations done to $c(r)$ can be summarized as the application of three ideas reported in the non-linear Optimized Random Phase Approximation (NL-ORPA) \cite{Pini2002} and the work of Sharma and Sharma (SS) \cite{Sharma1977}, which  combined are the core of this new proposal. These ideas are then used to obtain an analytical expression for $c(r)$, specifically for a SW system, leading us, through Eq. \eqref{eq: 02}, to an analytical expression for $S(k)$.

The first idea and simplification is contained in both NL-ORPA and SS approaches, and consists in the proposal of a direct correlation function that can be separated as the sum of two contributions
\begin{equation}
c(r) = c_0(r) + c_1(r),
\label{eq: 03}
\end{equation}
being $c_0(r)$ a reference contribution, directly associated with the hard-sphere (HS) interaction potential, while $c_1(r)$ is a perturbation term associated with the non-core part of the interaction potential. With such splitting in $c(r)$, the problem then translates on how to threat each of these two terms, which leads to the formulation of the remaining ideas.

In a similar manner as in the SS approach \cite{Sharma1977}, the reference part is approximated to be the Percus-Yevick solution of the HS system \cite{Wertheim1963}, with the distinction that Verlet-Weiss correction \cite{Verlet1972} is additionally implemented, thus the direct correlation function of the reference system is given by
\begin{equation}
c_0(r) =\left\{
\begin{array}{lcl}
\alpha + \delta r^* + \gamma {r^*}^3 & \hspace*{0.3cm} & r \leq \sigma \\
0 & \hspace*{0.3cm} & r > \sigma,
\end{array}
\right.
\label{eq: 04}
\end{equation}
where $r^*=r/\sigma$, $\alpha=-(1+2\phi)^2/(1-\phi)^4$, $\delta=6\phi(1+\frac{1}{2}\phi)^2/(1-\phi)^4$ and $\gamma=-\phi(1+2\phi)^2/2(1-\phi)^4$  are constants only dependent on the volume fraction  $\phi=\rho \sigma^3/6$, and in which the Verlet-Weiss correction takes the form of an empirical correction on the volume fraction $\phi_{VW}=\phi(1-\phi/16)$, due to an overestimation of the HS diameter $\sigma_{VW}=(\phi_{VW}/\phi)^{1/3}$. This proposal differs from ORPA-like schemes in a manner that it permits to obtain the full-functional form of $c(r)$, meanwhile in the ORPA-like schemes the problem in terms of $c(r)$ yields to numerically solve a set of equations to obtain $c_0(r)$ \cite{Khanpour2011}. It is important to mention that in this proposal, the volume exclusion property is not properly taken into account in the direct correlation function, in contrast with the ORPA-like schemes, nevertheless, and just as it is shown in the reference \cite{Sharma1977}, the results for the static structure factor, as well as for the radial distribution functions (for $r>\sigma$), have a good overall agreement when compared with a reference system.

The perturbation part is written in a similar manner than the NL-ORPA scheme proposal \cite{Pini2002}, as
\begin{equation}
c_{1}(r)=\left\lbrace
\begin{array}{lcl}
0 & \hspace*{0.3cm} & r \leq \sigma \\
e^{-\beta u(r)}-1 & \hspace*{0.3cm} & r > \sigma,
\end{array}
\right.
\label{eq: 05}
\end{equation}
where $\beta=1/k_B T$. We stress the fact that this functional form of $c_{1}(r)$ leads to an expected consistency for the small density expansion of $c(r)$, which is
\begin{equation}
c(r) \approx e^{-\beta u(r)}-1,
\label{eq: 06}
\end{equation}
a consistency that is also maintained within the proposal of $c_0(r)$, since  at small $\phi$ values $c_0\approx -1$.

With this two approximations done to both, $c_0(r)$ and $c_1(r)$, Eq. \eqref{eq: 04} and Eq. \eqref{eq: 05}, respectively, the direct correlation function, Eq. \eqref{eq: 03}, can be written for the SW fluid as
\begin{equation}
c(r) =\left\lbrace
\begin{array}{lcl}
\alpha + \delta r^* + \gamma {r^*}^3 & \hspace*{0.3cm} & r \leq \sigma \\
e^{1/T^*} - 1 & \hspace*{0.3cm} & \sigma < r \leq \lambda \sigma\\
0 & \hspace*{0.3cm} & r > \lambda\sigma,
\end{array}
\right.
\label{eq: 07}
\end{equation}
whose Fourier transformation lead us
\begin{eqnarray}
\hat{c}(k) & = & \hat{c}_{0}(k) \\
& + & \frac{4\pi x} {k^{*3}}\left[ k^{*} \left( \cos k^{*} - \lambda \cos \lambda k^{*} \right) + \left( \sin \lambda k^{*} - \sin k^{*} \right) \right], \nonumber
\label{eq: 08}
\end{eqnarray}
in which we define $x \equiv e^{1/T^*} - 1$, with $ T^{*}=1/\beta \epsilon $, $k^{*}=k\sigma$ is the reduced wave vector number, and $ \hat{c}_{0}(k) $ is given by
\begin{eqnarray}
\hat{c}_{0} \left( k^{*} \right) & = & \dfrac{4 \pi}{k^{*6}} \left\lbrace \alpha k^{*3} \left[ \sin k^{*} - k^{*}\cos k^{*} \right] + 24 \gamma \right.  \\
& + & \delta k^{*2}  \left[ 2k^{*} \sin k^{*} - \left(k^{*2} - 2 \right) \cos k^{*} - 2\right] \nonumber \\
& + & \left. \gamma \left[\left(k^{*3} - 24k^{*} \right) \sin k^{*} - \left( k^{*4} - 12k^{*2} + 24 \right) \cos k^{*} \right] \right\rbrace, \nonumber
\label{eq: 09}
\end{eqnarray}
then, using Eq. \eqref{eq: 08} we can explicitly write Eq. \eqref{eq: 02} for the structure factor as
\begin{eqnarray}
\label{eq: 10}
S \left( k \right) & = & \left[ 1 - \rho\hat{c}_{0}(k) \right. \\
& + & \left. \frac{4\pi \rho x }{k^{*3}} \left\lbrace  k^{*} \left( \cos k^{*} - \lambda \cos \lambda k^{*} \right) + \left( \sin \lambda k^{*} - \sin k^{*} \right) \right\rbrace \right]^{-1}.  \nonumber
\end{eqnarray}

The proposal for $c_{1}(r)$, given by Eq. \eqref{eq: 06} is the main difference when it is compared to the common ORPA scheme or the SS approach, both of them uses the high temperature limit of $c_{1}(r)=-\beta u(r)$, commonly referred as Random Phase Approximation (RPA), then, the above mentioned approaches can be recovered if a temperature expansion is performed over $ c_{1} \left( r \right) $, Eq. \eqref{eq: 05}, and the high temperature limit is taken into account, therefore $e^{1/T^*}\approx 1 + 1/T^*$, which yields the following couple of equations
\begin{eqnarray}
\hat{c}_{L}(k) & = & \hat{c}_{0}(k) \\
& + &  \frac{4\pi} {k^{*3}T^*}\left[ k^{*} \left( \cos k^{*} - \lambda \cos \lambda k^{*} \right) + \left( \sin \lambda k^{*} - \sin k^{*} \right) \right], \nonumber
\label{eq: 11}
\end{eqnarray}
and
\begin{eqnarray}
\label{eq: 12}
S_{L}(k) & = & \left[ 1 - \rho\hat{c}_{0}(k) \right. \\
& + & \left. \frac{4\pi \rho }{k^{*3}T^*} \left\lbrace k^{*} \left( \cos k^{*} - \lambda \cos \lambda k^{*} \right) + \left( \sin \lambda k^{*} - \sin k^{*} \right) \right\rbrace \right]^{-1},  \nonumber
\end{eqnarray}
which we referred as the linear version of our approach.

In fact, Eq. \eqref{eq: 12} is the same than the Eq. (5) derived by SS in Ref. \cite{Sharma1977}. Then, just as in the comparison between NL-ORPA and the common ORPA scheme, where the proposed functional form for $c_1(r)$ is a better overall estimation, the presented scheme is expected to be also a better estimation when it is compared to the SS scheme, this is at least easily proven in the low density regime since the functional temperature dependence is fully recovered as we show further bellow.

The relevance of a fully analytical expression for $ S\left( k \right) $ can be appreciated in the determination of thermodynamic variables, for example, by means of the relationship between $ S\left( k \right) $ and the isothermal compressibility one can obtain the pressure equation of state
\begin{equation}
P^{*} = \dfrac{6T^{*}\phi}{\pi} \left[ \frac{\left( 1+\phi + \phi^2 -\phi^3 \right)}{ (1-\phi)^3 }\right] - \dfrac{24}{\pi} \left( \lambda^{3} -1 \right) T^{*} x \phi^{2},
\label{eq: 13}
\end{equation}
where $P^*=\sigma ^3 P/\epsilon$ is the reduced pressure, furthermore, if the linear version of $ S \left( k \right) $ is taken into account the obtained pressure equation is
\begin{equation}
P^{*} = \dfrac{6T^{*}\phi}{\pi} \left[ \frac{\left( 1+\phi + \phi^2 -\phi^3 \right)}{ (1-\phi)^3 }\right] - \dfrac{24}{\pi} \left( \lambda^{3} -1 \right) \phi^{2}.
\label{eq: 14}
\end{equation}

A detailed derivation of the above equations can be found in the appendix \ref{App1}. Since a thermodynamic consistency between different routes to determine the pressure equation it is not expected, a one-to-one comparison of this quantity should be employ the same route, a detailed revision of this point can be found in sec. \ref{SecV}.

\section{\label{SecIV}Microscopical structure of the SW fluid}

\begin{figure}[htp!]
	\includegraphics[scale=0.49]{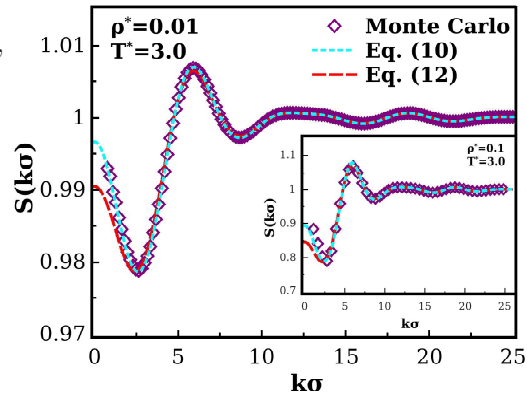}
	\caption{Structure factor at high temperature $ T^{*}=3.0 $ and density $ \rho^{*}=0.01 $ and (inset) $ \rho^{*}=0.1 $ for the $ \lambda = 1.5 $ SW fluid. Results from (diamonds) Monte Carlo simulations in the $ NVT $ ensemble, (red dashed line) linear or SS approach and (cyan dotted line) Eq. \eqref{eq: 10} of this work.}
	\label{Fig: 03}
\end{figure}

Let us now proceed to demonstrate, in the low density regime, how the analytical expression given in Eq. \eqref{eq: 10} is more accurate than its linear form, which is the SS approach, see Eq. (5) in reference \cite{Sharma1977}. As is discussed above, the main difference between our proposal and the SS approach resides in the temperature dependence of the direct correlation function, which is expected to reproduce the expected physical behaviour at the low density limit, see Eq. \eqref{eq: 06}. Thus, in this limit one could expect a high degree of accuracy respect to the exact results, in this study we take the MC simulation results as a reference exact results, the explored thermodynamic states with our theoretical approach are shown in Fig. \ref{Fig: 01}. For instance, in Fig. \ref{Fig: 03}, we compare the microscopical structure predicted by the theoretical approaches, dashed line and dotted line for the linear and full Eq. \eqref{eq: 10}, respectively, against the MC results shown with diamonds.

For $ k \sigma > 2 \pi $, the agreement between both theoretical approaches is high respect to the simulations. Nevertheless at lower $ k \sigma $ values and despite that in this particular case a moderate high temperature is analysed the linear form of Eq. \eqref{eq: 10} fails to reproduce the structure factor at such wave vectors, furthermore, this behaviour is independent of the density as one can see in the inset of Fig. \ref{Fig: 03}. A similar analysis was performed (data not shown) for different temperatures and $ \lambda $ and the trend described before is the same, hence, our approach retains its good agreement respect to the MC results even if the density is increased.

Another important property to characterize the microscopical structure is the radial distribution function $ g(r) $ \cite{Hansen}. This quantity can be directly obtained performing a FT on the $ S\left( k \right) $ as
\begin{equation}
g\left( r \right) = 1 + \dfrac{1}{\rho}\int_{{\bf V}} d{\bf r} \; \left[ S(k) - 1 \right] e^{i {\bf k r}}.
\label{eq: 15}
\end{equation}

\begin{figure}[htp!]
	\includegraphics[scale=0.49]{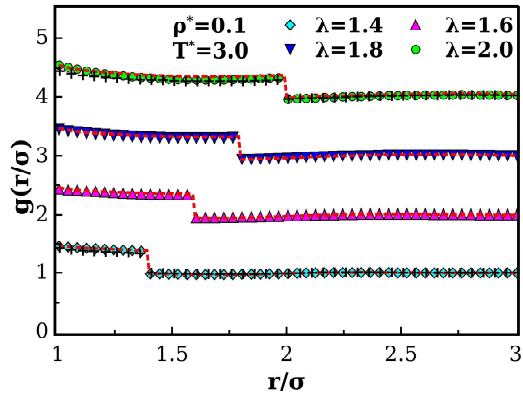}
	\caption{Radial distribution function for different SW fluids at density $ \rho^{*}=0.1 $ and temperature $ T^{*}=3.0 $. The symbols are MC results and dashed lines are the predictions of Eq. \eqref{eq: 11} with the full temperature dependence and the (black) crosses are the results of its linear version.}
	\label{Fig: 04}
\end{figure}

Employing Eq. \eqref{eq: 15}, first we determine $ g\left( r \right) $ for different values of $ \lambda $ at several thermodynamic states, hence, in Fig. \ref{Fig: 04} we only show results for selected values of $ \lambda $ at $ \rho^{*}=0.1 $ and $ T^{*}=3.0 $, see Fig. \ref{Fig: 01} to identify such thermodynamic state in their respective phase diagram. The symbols in the figure stands for the MC results, the red dashed line are the predictions of Eq. \eqref{eq: 15} and the black crosses are results of its linear version, it means, the high temperature limit of Eq. \eqref{eq: 10} has been taken into account to compute Eq. \eqref{eq: 15}, for the sake of clarity, in the last case, only results for $ \lambda = 1.4 $ and $ \lambda = 2.0 $ are shown.

\begin{figure}[htp!]
	\includegraphics[scale=0.49]{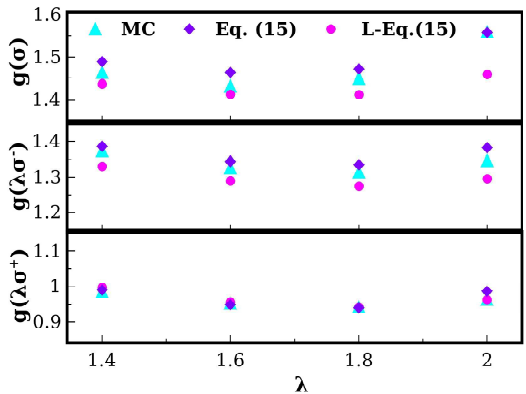}
	\caption{Contact and discontinuity values of the radial distribution function as a function of $ \lambda $. The triangles, circles and diamonds are results of MC, Eq. \eqref{eq: 15} and its linear form, respectively. The thermodynamic state analysed is the same than the one in the main frame of Fig. \ref{Fig: 03}.}
	\label{Fig: 05}
\end{figure}

At first sight one can claim the existence of a very good agreement between theoretical and simulations results, nevertheless, a close inspection reveals to us that the values of $ g(r) $ at the contact and the first discontinuity $ r = \lambda \sigma^{-} $ are slightly overestimated as the $ \lambda $ value increase, see for instance Fig. \ref{Fig: 05}, besides the linear version of Eq. \eqref{eq: 15} underestimate both the contact and discontinuity values of $ g(r) $.

\begin{figure}[htp!]
	\includegraphics[scale=0.49]{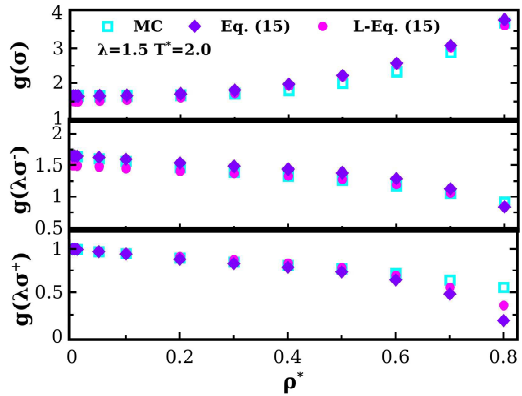}
	\caption{Contact and discontinuity values of the radial distribution function as a function of $ \rho^{*} $ for $ \lambda=1.5 $ SW fluid at $ T^{*}=2.0 $. The squares, diamonds and circles are results of MC, Eq. \eqref{eq: 15} and its linear form, respectively.}
	\label{Fig: 06}
\end{figure}

On the other hand, the density dependence of the contact and discontinuity values of $ g \left( r \right) $ shows a good overall behaviour as one can see in Fig. \ref{Fig: 06}, where we show results for $ g\left( \sigma \right) $, $ g\left( \lambda \sigma^{-} \right) $ and $ g\left( \lambda \sigma^{+} \right) $ for thermodynamic states where the temperature is fixed at $ T^{*}=2 $, see Fig. \ref{Fig: 02} for a visual reference of those thermodynamic states. Even at this high temperature the structure given by Eq. \eqref{eq: 15} offers better results at low densities than its linear version, although, both theoretical approaches has deviations respect to MC results for densities higher than $ \rho^{*}=0.4 $, for smaller values, Eq. \eqref{eq: 15} predicts almost perfectly the MC results. It is worth to mention that the deviations observed at high densities can be due to a volume exclusion property which is not properly taken into account in our approach, nevertheless, such differences are small, see further bellow the discussion around Fig. \ref{Fig: 09} at this respect.

We stress the fact that in the case of the SW fluid the contact and discontinuity values of the radial distribution function are of great interest since, as we discuss further bellow, this information is needed to determine the pressure equation. Thus, in this way the determination of the pressure equation completely depends on the computation of such quantities.

\begin{figure}[htp!]
	\includegraphics[scale=0.49]{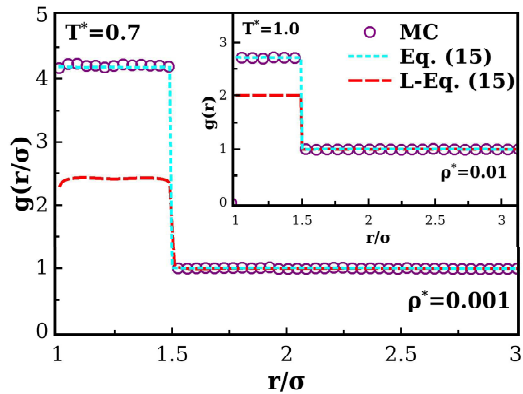}
	\caption{Radial distribution function at the low density regime, namely $ \rho^{*}=0.001 $ at $ T^{*}=0.7 $ in the main frame and $ \rho^{*}=0.01 $ at $ T^{*}=1.0 $ in the inset. Circles, dotted and dashed lines stands for MC, Eq. \eqref{eq: 15} and its linear form results, respectively.}
	\label{Fig: 07}
\end{figure}

Now, let us focus in the low density regime, which again we characterize through the radial distribution function, therefore in Fig. \ref{Fig: 07} we analyse the results predicted in the low density regime, namely, $ \rho^{*} =0.001$ and $ \rho^{*}=0.01 $, also we consider temperatures bellow the critical one of the $ \lambda=1.5 $ SW fluid. Such thermodynamic states are in the homogeneous phase of the fluid as one can see in Fig. \ref{Fig: 02}. In this scenario, our approach, given by Eq. \eqref{eq: 10} in combination with Eq. \eqref{eq: 15}, shown as a dotted line, displays a high degree of accuracy respect to the MC results, however, the linear form of our approach (dashed line), has a very poor performance, in such a way that in the range $  \sigma \leq r \leq \lambda \sigma $ the radial distribution function is underestimated, then, even a qualitative comparison is forbidden. It is at this regime of low density and temperature where our approach shows its relevance since it outperform the SS analytical proposal for the computation of the structure, which we remember can be obtained as a limit case of the proposal of this work and we called linear version.

In general terms, our approach slightly overestimates both $ g (\lambda \sigma ^{-} ) $ and $ g (\lambda \sigma ^{-} ) $, but, works very well in the regime for which it was designed even at low temperatures.  Nevertheless, we know there are alternately theoretical schemes to compute the structure of SW fluids, like the ones based on the integral equations formalism, for example, however, such methods needs iterative algorithms of solution, implying complex numerical implementations with potential problems of convergence. In contrast, the Eq. \eqref{eq: 10} is a closed and fully analytic expression, with a straightforward implementation that works very well in a wide range of densities and temperatures. Besides, is exact at the low density limit.

\section{\label{SecV}Equation of state: A test of accuracy}
Encouraged by the above results we employ the microscopical structure to determine another thermodynamic properties, since from $ g\left( r \right) $ one can compute the system total energy, pressure and the chemical potential \cite{Allen1991}, we decide to compute the pressure equation of state and with that establish the degree of accuracy of our approach. The route to compute the pressure through $ g\left( r \right) $ is given by
\begin{equation}
PV = Nk_{B}T - \dfrac{2}{3}\pi N \rho \int_{0}^{\infty} r^{2} \dfrac{d u \left( r \right)}{d r} g \left( r \right) dr.
\label{eq: 16}
\end{equation}

However, since the SW interaction potential and the $ g(r) $ have discontinuities the Eq. \eqref{eq: 16} can not be computed straightforwardly. Smith and co-workers \cite{Smith1977} proposed an equation that use only the contact and discontinuities values of $ g \left( r \right) $ whose relationship with the pressure is given by
\begin{eqnarray}
P^{*} & = & \rho^{*}T^{*} + \dfrac{2\pi}{3} \rho^{*2} T^{*} g \left(\sigma^{+} \right) \nonumber \\
& - & \dfrac{\pi}{3} \rho^{*2} \lambda^{3} \left[ g \left( \lambda \sigma^{-} \right) + g \left( \lambda \sigma^{+} \right) \right],
\label{eq: 17}
\end{eqnarray}
where $ g \left(\sigma^{+} \right) $ is the contact value, $ g \left( \lambda \sigma^{-} \right)$  and $ g \left( \lambda \sigma^{+} \right) $ are the discontinuities values of the radial distribution function at $ \lambda \sigma $, respectively. The Eq. \eqref{eq: 17} is one of the reasons for which the contact and discontinuities values of $ g(r) $ are of great interest. From theoretical point of view, there are different routes to compute the pressure, as we already mentioned, one of them is the route of the isothermal compressibility, which in our case gives us the Eq. \eqref{eq: 14}, whose derivation details can be seen in the appendix \ref{App1}. Nevertheless, in order to made a systematic analysis, for the time being, we use Eq. \eqref{eq: 17} for both, theoretical and simulation results.\\

\begin{figure}[htp!]
	\includegraphics[scale=0.52]{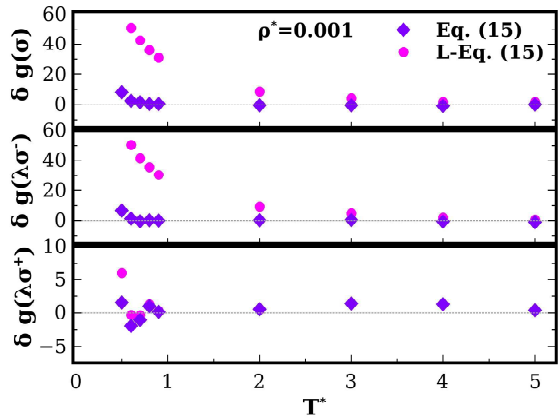}
	\caption{Percentage relative errors (from top to bottom) of the contact $ g \left( \sigma \right) $ and discontinuity values $ g \left( \lambda \sigma^{+} \right) $, $ g \left( \lambda \sigma^{-} \right) $ as a function of the temperature for a $ \lambda = 1.5 $ SW fluid. Diamonds and circles are results of Eq. \eqref{eq: 11} and its linear version. The line is just a reference to the eye.}
	\label{Fig: 08}
\end{figure}

\begin{figure}[htp!]
	\includegraphics[scale=0.52]{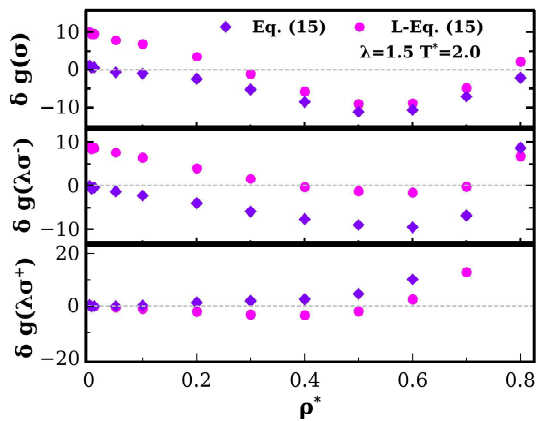}
	\caption{Percentage relative errors (from top to bottom) of the contact $ g \left( \sigma \right) $ and discontinuity values $ g \left( \lambda \sigma^{+} \right) $, $ g \left( \lambda \sigma^{-} \right) $ as a function of the density for the same conditions that in Fig. \ref{Fig: 06}. Diamonds and circles are results of Eq. \eqref{eq: 11} and its linear version, respectively.}
	\label{Fig: 09}
\end{figure}

As is discussed above respect to the structure, the overall agreement of our approach with the simulation results is good, nevertheless, if such differences are quantified could give us an idea of the expected predictions for the pressure. Thus in Fig. \ref{Fig: 08} and Fig. \ref{Fig: 09} we show a couple of representative cases of the error for the contact and discontinuity values of $g(r)$ between our approach and MC results as a function of temperature and density, respectively. We define the percentage relative error as
\begin{equation}
\delta R \equiv 100 \times \dfrac{R_{MC} - R_{T}}{R_{MC}},
\label{eq: 18}
\end{equation}
where $ R $	stands for a particular result, i.e., any $ g \left( \sigma \right) $, $ g \left( \lambda \sigma^{-} \right) $ or $ g \left( \lambda \sigma^{+} \right) $, the subscript $ MC $ and $ T $ means Monte Carlo and theoretical results. Then a positive value of this quantity indicates an underestimation and a negative one an overestimation. Hence, diamonds and circles in referred figures stands for results of Eq. \eqref{eq: 15} and its linear form, respectively.

From Fig. \ref{Fig: 08} it is clear that at high temperatures, $ T^{*} \geq 3 $, in the low density regime, both versions of our approach have almost the same deviation, which is lesser than $ 5\% $. Nevertheless, as the temperature is decreased below $ T^{*}=1 $ the linear version of our approach has deviations greater than the $ 20\% $, being $ g \left( \sigma \right) $ and $ g \left( \lambda \sigma^{-} \right) $ the quantities with such error. On the other hand, if the same errors are analysed as a function of the density, see Fig. \ref{Fig: 09}, our approach gives excellent results for densities lower than $ \rho^{*}=0.2 $ since its error is lesser than $ 10 \% $, however as the density increase beyond than $ \rho^{*}=0.4 $ both versions of our approach has errors greater than $ 10 \% $ being the contact value overestimated in this regime.

\begin{figure}[htp!]
	\includegraphics[scale=0.49]{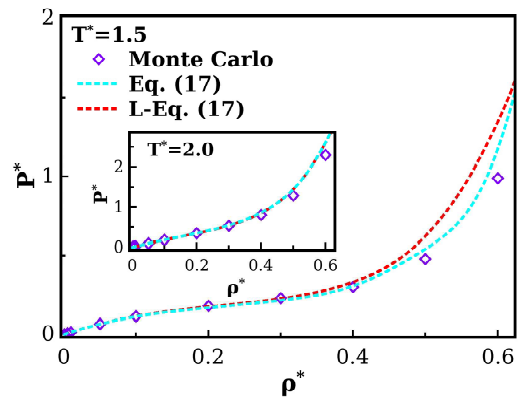}
	\caption{Pressure equation of state as a function of density for the $ \lambda=1.5 $ SW fluid at $ T^{*}=1.5 $ in the main frame and $ T^{*}=2.0 $ in the inset. Symbols stands for Monte Carlo simulations in the NVT ensemble whereas (cyan) dotted line and (red) dashed line are the theoretical results predicted by Eq. \eqref{eq: 17} and its linear version, respectively.}
	\label{Fig: 10}
\end{figure}

Now, turning our attention to the pressure equation, in Fig. \ref{Fig: 10} results for the $ \lambda = 1.5 $ SW fluid at $ T^{*}=1.5 $ and $ T^{*}=2.0 $ are shown as a function of the reduced density. The MC results are represented by diamonds and the theoretical ones are shown with dotted and dashed lines for the Eq. \eqref{eq: 17} and its linear version, respectively. At $ T^{*}=2 $, the theoretical results are almost the same, although respect to the MC results differences can be glimpsed at reduced densities higher than $ \rho^{*} > 0.4 $. A similar trend is found if the temperature is decreased at $ T^{*}=1.5 $, but in this case the deviations are more easily seen and results in an overestimation of both versions of our approach, however, until $ \rho^{*}=0.6 $ the Eq. \eqref{eq: 17} gives better results than its linear version.

From Fig. \ref{Fig: 10} one can see that the qualitative agreement between both versions of our theoretical approach and the simulation results is very good, despite that the linear version of our approach has errors greater than $ 10\% $ the agreement with MC results is good since as one can see in Eq. \eqref{eq: 17} the contact and discontinuity values of $ g(r) $ are multiplied by a squared factor of $ \rho $, then at low densities such contributions are small. In general terms, the performance of Eq. \eqref{eq: 10} and Eq. \eqref{eq: 15} as the quantities derived from it with the full temperature dependence are better than its linear version, furthermore, we can confirm that in the low density regime and also low temperatures our approach is far better than its linear version.

\section{\label{SecVI}Concluding remarks}
We present an equation for the static structure factor based on the non-linear ORPA and SS ideas while proposing the usage of the analytical solution of the HS system for the reference part of $c(r)$, therefore making possible an analytic expression. Within this proposal, as it is expected, obeys the high temperature limit for all densities (including the HS limit found at infinite temperature), and the small densities regime. The structure is found to be, for all the studied cases, a better approximation than the linear version when compared with MC simulations. The pressure equation of state derived from the theoretical approach is also found to have a very good agreement with results from MC simulation in a wide range of densities and different temperatures, which just as the structure, outperform the results obtained by the linear version. Despite the existence of more robust approaches to determine the structure or the thermodynamics properties of the SW fluid, the obtained analytic expression to compute the structure factor is easy to implement and practically has no computational cost, making it a versatile tool for studies that requires systematic analysis.

The structure then is expected to have good results when used as an equilibrium input on the existent dynamical diffusion theories such the Enskog \cite{Higgins1958,Davis1961}, MCT\cite{Janssen2018,Goetze1998} and SCGLE\cite{Maldonado2007,Ramirez2010} theories. Additionally, it is expected to outperform the linear version of this structure on the description of non-equilibrium conditions in the NE-SCGLE theory at the small density regime, which has already been used in the framework of a Yukawa perturbation potential \cite{Olais2015}. Lastly, the framework described in the third section is expected to work for a variety of small range interaction potentials, such as the already mentioned Yukawa potential, although the expressions in general would be denoted in terms of the Fourier transform of the perturbation, which is not necessarily an analytic transformation.

\section*{Acknowledgements}
The  authors  thankfully  acknowledge  computer  resources,  technical advise and  support provided  by  Laboratorio  Nacional  de  Supercómputo  del  Sureste  de  México (LNS), a  member of the CONACYT national laboratories, with project No. 201901035N. A.T.C. and P.E.R.G. acknowledge the financial support of CONACyT through grants: Estancias Postdoctorales Nacionales No. 422753/2018  and C\'atedras CONACyT No. 1631 and CB-2015-01-257636. The authors would like to thank the national laboratory LANIMFE for the infrastructure provided during this project.

\appendix

\section{\label{App1}Pressure equation of state for square well fluid}
The equation of state of the Square Well system can be obtained through the relation between the system structure factor and the isothermal compressibility. This relation is stated as follows:
\begin{equation}
\lim_{ k \to 0 } S(k) = \chi_T^*,
\end{equation}
where $\chi_T^*=\chi_T/\chi_T^{id}$ is the system isothermal compressibility divided by the ideal gas isothermal compressibility. With this expression, the isothermal compressibility of the SW system in this work can be expressed as:
\begin{equation}
\chi_T^* = \frac{1}{\chi_{T,HS}^{*-1} - 8\phi \left( \lambda^3 - 1 \right) x },
\end{equation}
where $x= e^{1/T^*}-1$, and where $\chi_{T,HS}^{*}$ is the obtained isothermal compressibility of the reference system, which with the Percus-Yevick approximation and with Verlet-Weiss correction is in a good agreement with the Carnahan-Starling compressibility\cite{Verlet1972}:
\begin{equation}
\chi_{T,HS}^{*} = \frac{8\phi - 2\phi^2}{ (1-\phi)^4 }.
\end{equation}
Through the definition of the isothermal compressibility, which can be stated in terms of the volume fraction as:
\begin{equation}
\chi_T \equiv \frac{1}{\phi} \left(  \frac{\partial \phi}{\partial P} \right)_T,
\end{equation}
the equation to solve for $P$ can be written as:
\begin{equation}
\frac{\partial P}{\partial \phi} = \left( \frac{\partial P_{HS}}{\partial \phi} \right) - 8 \left( \lambda^3 - 1 \right) \phi T^* x,
\end{equation}
where $P_{HS}$ is the pressure for the reference system:
\begin{equation}
\frac{P_{HS}}{\rho k_B T} = \frac{1+\phi + \phi^2 -\phi^3}{1-\phi}^3,
\end{equation}
and whose solution is given by:
\begin{equation}
P^* = P^*_{HS} - \frac{24}{\pi} \left( \lambda^3 - 1 \right)T^* x \phi^2,
\end{equation}
where $P^*=\sigma^3P/\epsilon$ and  $P^*_{HS}= \sigma^3 P_{HS}/ \epsilon$. From this equation, the equivalent equation state from SS approximation can be obtained through the high temperature limit, in which $x\approx 1/T$:
\begin{equation}
P^*_{SS} = P^*_{HS} - \frac{24}{\pi} \left( \lambda^3 - 1 \right) \phi^2.
\end{equation}

\bibliography{Square.bib}

\end{document}